\documentstyle[epsf,editedvolume]{crckapb}

\begin{opening}
\title{Globular Cluster Systems of Elliptical Galaxies}

\author{Stephen E. Zepf}
\institute{Dept.\ of Astronomy, Yale University, New Haven, CT 06520}
\author{Keith M. Ashman}
\institute{Depts.\ of Physics, Univ.\ of Kansas, Lawrence, KS, 66045 \\
and Baker University, Baldwin, KS 66006}

\end{opening}

\runningtitle{Globular Cluster Systems}

\begin{document}


\begin{abstract}

\vskip 2pt
We review the observed properties of globular
cluster systems and their implications for models of
galaxy formation. Observations show that globular clusters
form in gas-rich mergers,  and that bimodal metallicity distributions
are common in the globular cluster systems of ellipticals, with the
metal-poor population more extended than the metal-rich one.
These are three of the four predictions of the simple merger
model of Ashman \& Zepf (1992). The fourth prediction concerns
the properties of the globular cluster systems of spirals, and is
still to be tested by observation.  Adopting Occam's
razor, the confirmation of the fundamental predictions of the merger
model from both young and old globular cluster systems is strong evidence that
typical elliptical galaxies formed from the mergers of spiral galaxies.
However, the simplifying assumptions of the Ashman-Zepf
merger model limit its applicability to certain complex situations such
as the formation of cD galaxies. We conclude this review
by introducing new observational and theoretical
programs that will further  the understanding of the physical
mechanisms of globular cluster and galaxy formation.

\end{abstract}

\vspace*{-18pt}

\section{Introduction}

The dramatic revolution in the understanding
of globular cluster systems and their implications for galaxy formation
can be traced by following the role
the subject has played at major international meetings on
mergers of galaxies. At the Heidelberg meeting in 1989,
globular cluster systems (GCSs) were the focus of the ``appointed
 skeptic''  (van den Bergh 1990), whereas in Kyoto in 1997,
globular clusters appeared prominently in the introductory talk
as some of the strongest evidence that now quiescent elliptical
galaxies formed from the past mergers of spiral galaxies
(Schweizer 1998). This review will discuss the theoretical
predictions and observational evidence that led to this revolution.
We will then point to new directions that promise further
advances in our understanding of globular cluster systems
and their implications for the formation history of their
host galaxy.

\vspace{-6pt}

\section {What the Merger Model Predicted}

\vskip -2pt
Ashman and Zepf (1992; hereafter AZ) carried out a detailed
study of the relationship between
galaxy mergers and globular cluster systems.
If elliptical galaxies are the products of spiral galaxy mergers, globular
clusters must form in such mergers (Schweizer 1987). This is because
elliptical galaxies have a higher specific frequency (number per unit
luminosity) of globular clusters than do spirals (van den Bergh 1990).
By considering globular cluster formation efficiency and the gas content
of the progenitor spiral galaxies, AZ showed that a sufficient number
of globular clusters could form in major mergers to explain the specific
frequency of globular clusters systems around ellipticals.
This led to the conceptually simple prediction that, if elliptical galaxies
are formed in galaxy mergers, ongoing galaxy mergers contain
young globular clusters. In addition, AZ described the expected properties
of young globular clusters, such as their blue colors, bright
luminosities, and compact sizes. They also pointed out that HST observations
of ongoing mergers like NGC~7252 and NGC~1275 should clearly
reveal such objects.
Thus the presence of young
globular clusters in ongoing mergers constitutes a testable prediction
that can be used to {\it refute} the merger model.

	Perhaps the most unique prediction of the Ashman-Zepf
model was that the globular cluster systems of elliptical galaxies 
formed in mergers have bimodal metallicity distributions.
The progenitor spirals are expected to have
a halo of metal-poor clusters, like those observed in the Milky Way and M31.
More generally, in the context of the simple merger picture one expects
such globular clusters to have formed out of gas that has experienced little
metal enrichment. In contrast, globular clusters that form in the merger itself
are expected to form out of the relatively enriched gas in the spiral disks.
Thus these ``merger-produced'' globular clusters will have significantly
higher metallicities than their counterparts in the halos of the progenitor
spirals.  For elliptical galaxies with ``normal'' specific frequencies
(roughly double those of spirals), AZ showed that the number of
metal-rich globular clusters formed in the merger must be comparable to
the number of metal-poor globular clusters contributed by the progenitor
spirals.  Thus the overall metallicity distribution of the globular cluster
systems of ellipticals is predicted to be bimodal, with roughly equal
numbers of clusters in each ``peak'' of the distribution.

The third prediction of the Ashman-Zepf model is that, in an elliptical
galaxy, the metal-rich cluster system
will be more centrally concentrated than the metal-poor cluster system.
This difference mirrors the different histories of the gas out of which
the two globular cluster subsystems form.
Compared to the low-metallicity halo gas that produces the metal-poor
globular clusters, the enriched gas responsible for the metal-rich clusters
is likely to have undergone more dissipation. This may arise as gas collapses
from the halo to the disk in the progenitor spirals, or during the merger
process itself. Consequently, the metal-rich cluster system is predicted
to be more spatially concentrated than the metal-poor system. The effect may
be increased by the tendency of mergers to ``puff up'' the stellar
components of the progenitor galaxies. This would affect the metal-poor
globular clusters, but not the dissipative gas out of which the metal-rich
clusters form.

\vspace{-6pt}
\section {What the Observations Showed}

\vskip -2pt
        Observational studies of globular cluster systems have
advanced rapidly, so the fundamental predictions of the Ashman-Zepf model
can now be tested. This section is devoted to a review of the
relevant observations and the comparison of these to the model
predictions.

\vspace{-8pt}
\subsection {Young Globular Clusters}

\vskip -4pt
	The prediction that globular clusters form in gas-rich mergers has
now been repeatedly confirmed, as reviewed in many places
(e.g. Whitmore, these proceedings, Schweizer 1997, Ashman \& Zepf 1998).
Specifically, HST observations of gas-rich mergers have uncovered 
a wealth of objects with bright luminosities, blue colors, and 
compact sizes, as predicted by AZ. These properties are all consistent 
with those expected of young objects with masses and sizes
of the globular clusters in the Galaxy. Moderate resolution spectroscopy
has further confirmed that these objects are composed of a young
stellar population consistent with those given by standard stellar
population models.
This agreement includes the strength of the Balmer lines, when compared
to LMC clusters (Bica \& Alloin 1986) or to updated
stellar populations models and their improved stellar libraries are
used, which were not available when the first spectra were published
(e.g.\ Schweizer \& Seitzer 1993, Zepf et al.\ 1995a).
Finally, high resolution spectroscopy of a few of the most nearby
examples provides velocity dispersions indicative of masses
that are typical of Galactic globular clusters, and agree well 
with those calculated from the observed colors and luminosities, 
combined with stellar population models (Ho \& Fillipenko 1997).

        The prediction that globular clusters are formed in merger-induced
starbursts is therefore confirmed. However, it is also important to 
recognize that well established physical processes will act to destroy 
some fraction of the initial young globular clusters (Fall \& Rees 1977
and subsequent papers). 
Such destruction is also suggested by the higher ratio of light
in clusters to total light seen in galaxy mergers compared to any old
system, even the halo of the Galaxy or cD galaxies (Zepf et al.\ 1998).
 Moreover,
the net effect of these processes is to preferentially destroy low mass
clusters, so it is possible to begin with a power-law mass function,
and end-up with a lognormal mass function, like that observed in the galaxy
(e.g.\ Gnedin \& Ostriker 1997).

\vspace{-6pt}
\subsection{Bimodal Color Distributions}

\vskip -2pt
 The second prediction of the AZ merger model is that the globular
cluster systems (GCSs) of elliptical galaxies will be composed of
metal-poor populations from the progenitor spirals and metal-rich
populations formed during the merger(s) that formed the elliptical.
As described by Zepf \& Ashman (1993) and Ashman \& Zepf (1998), the
metal-rich population will typically be significantly redder
than the metal-poor population because most mergers occur at
moderate or high redshift, and therefore metallicity differences
dominate the broad-band colors.

        Bimodality in the color distribution of the GGCs of elliptical 
galaxies was discovered by Zepf \& Ashman (1993).
Using the best available data and mixture-modeling algorithms,
they showed that the color distributions of NGC~4472 and NGC~5128
were likely to be bimodal.
Data of much higher quality for a number of elliptical galaxies
have since become available. This large body of evidence indicates
that bimodality is the norm for bright ellipticals (e.g.\ Ashman \& Zepf 1998).
Two typical examples are shown in Figure 1. In these and
many other cases, mixture-modeling algorithms confirm objectively
the significance of the bimodality apparent to the eye. 
The reader is referred to the Ashman \& Zepf (1998) book 
for plots of many more examples of color distributions of GCSs. 

\begin{figure}
\vspace{-140pt}
\epsfysize=4.2in
\hskip 0.75in
\epsffile{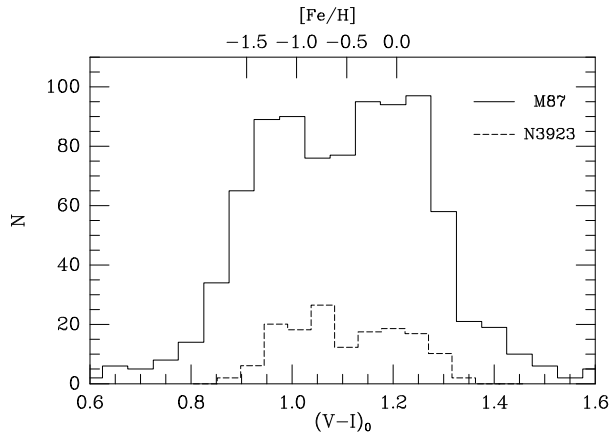}
\caption{Color distributions for the GCSs of M87 from Whitmore et al.\ (1995)
and NGC~3923 from Zepf et al.\ (1995b). The color distributions of both GCSs
are bimodal, although the number of clusters observed in M87 is much greater}
\end{figure}

\vspace{-6pt}
\subsection {Color Gradients}

\vskip -2pt
        The third prediction of the AZ model is that the
the metal-rich population is more spatially concentrated than
the metal-poor population. Color gradients, which were
suspected at the time of AZ and have now been confirmed
(Ashman \& Zepf 1998 and references therein), are a natural
result of the AZ model, although not unique to it.
The key to testing the merger prediction is to 
compare directly the distributions of the blue and red populations
around elliptical galaxies.
This was first achieved by the Geisler et al.\ (1996)
study of the NGC~4472 system. They showed that the
metal-rich population is more spatially concentrated than
the metal-poor population, thereby confirming the third 
prediction of the AZ merger model.

\vspace{-6pt}
\subsection {Other Considerations}

\vskip -2pt
While the fundamental predictions of the Ashman-Zepf model have
been confirmed, there are specific cases where the quantitative
agreement between prediction and observation breaks down.
This is currently most clear for the number and precise color of the
metal-poor population in elliptical galaxies 
(e.g.\ Zepf et al.\ 1995b, Forbes et al.\ 1997).
In the simplest merger picture, the color and specific frequency ($S_N$)
of the blue population is constant because all progenitor spirals are 
assumed to have the same halo population of metal-poor clusters. 
Higher specific frequencies of clusters in some ellipticals are 
therefore attributed to increased formation (or survival) efficiency 
of metal-rich clusters created in the mergers that made these galaxies. 
However, the metal-poor globular cluster populations
of ellipticals do not appear to have identical metallicities, and
some high $S_N$ systems have an enhanced frequency of metal-poor
clusters. Perhaps the clearest case is M87, for which the specific
frequency of metal-poor clusters is about seven: much higher than
the observed specific frequency of metal-poor globulars in spirals.
However, this may not be universal, as the high $S_N$ of
NGC~3311 appears to be due solely to metal-rich clusters (Secker et al.\ 1995).

	It is clear that these observations indicate the breakdown
of the simplifying assumptions of the model. Brighter ellipticals
are unlikely to be the result of the merger of only two spirals;
they are too massive. Further, progenitor spirals will not have identical
cluster populations before the merger. Finally, accretion of smaller
satellites is likely to play a role as well, as seems to be the case
for the halo population of the Galaxy. The question is whether a more
sophisticated and realistic treatment of the merger process is likely
to preserve these two fundamental predictions of the Ashman-Zepf model.
While a complete answer to this question requires detailed modeling 
(see Section 4), it seems inevitable that metallicity bimodality and 
the spatial concentration
dichotomy {\it will} be preserved, at least in some ellipticals. This is
because the merger model requires that major spiral-plus-spiral mergers
are involved at some point in the formation of history of an elliptical.
The number of such mergers and the likelihood that later mergers may be
predominantly stellar are both unimportant, since {\it typically} globular
clusters in the progenitor spiral halos will be metal-poor, and those
formed in any gas-rich merger will be
relatively metal-rich and spatially concentrated.

\section {New Observational and Theoretical Paths}

\vspace{-4pt}
\subsection {Theory}

\vskip -2pt
One way to advance the model beyond the formation
of an elliptical from the merger of two spirals	is to place
the model in a more specific cosmological context. In this way,
the full merging history of elliptical galaxies can be followed
statistically for a variety of possible models. Semi-analytic
models of galaxy formation and evolution (e.g.\ Kauffmann et al.\ 1993,
Cole et al.\ 1994) are well-suited for this program, and one
of us (SEZ) is working with G. Kauffmann to implement this approach.
A second area ripe for advancement is the theoretical understanding
of the formation of globular clusters, as the wealth of new observational
evidence in nearby merging galaxies has not yet been turned in to advances
in the understanding of the formation of dense, bound stellar systems
like globular clusters. 

\vspace{-6pt}
\subsection {Observation}

\vskip -2pt
	The kinematics of globular cluster systems provides
valuable information about the formation history and mass
distribution of the host galaxy. Until recently, it
had proven to be difficult to obtain more than a few tens of
clusters around a given galaxy. However, the increase in efficiency
and areal coverage of multi-object spectrographs, as well as higher
quality imaging for object selection, has opened up this field,
even with 4-m class telescopes.

\begin{figure}
\epsfysize=4.3in
\hskip 0.75in
\epsffile{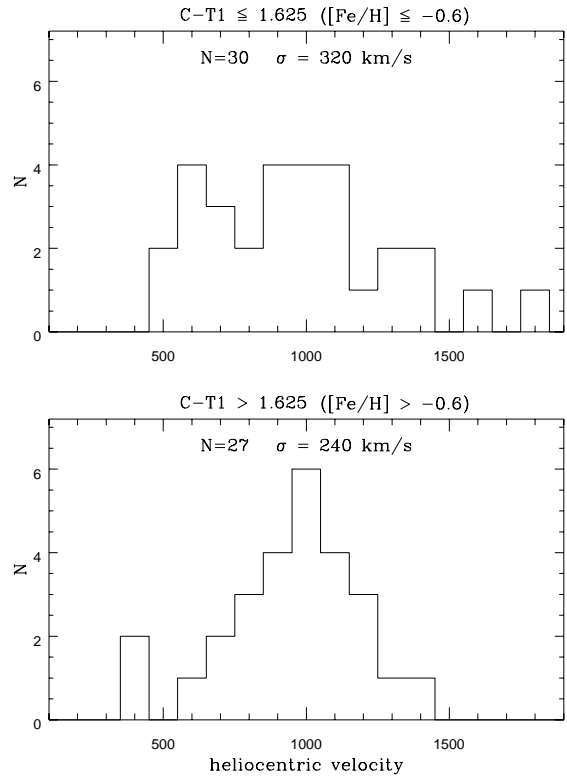}
\vspace{-18pt}
\caption{A comparison of the velocity dispersions of the metal-poor
and metal-rich cluster populations in NGC~4472. An F-test rejects
the hypothesis that these two have the same dispersion at the $86\%$
confidence level.}
\vspace{-16pt}
\end{figure}

	One of the best studied cases is M87 (Cohen \& Ryzhov 1997
and references therein).
Here, we will focus on our work with R. Sharples and others on 
spectroscopy of globular clusters around NGC~4472 (Sharples et al.\ 1997). 
The most exciting result to come from our study
is the tentative detection of kinematical differences between the
metal-rich and metal-poor cluster populations in this galaxy.
Specifically, the velocity dispersion of the metal-rich clusters
appears to be higher than that of the metal-poor clusters (Figure 2). 
Moreover, the metal-poor clusters appear to rotate along the major axis,
whereas the metal-rich clusters do not. At a basic level, the 
kinematics provide further physical evidence for the division 
of the cluster system into metal-poor and metal-rich subsystems
that was originally based on an analysis of the colors alone.
Furthermore, since the metal-rich cluster population is more 
spatially concentrated but has less rotation than the metal-poor 
cluster system, significant angular momentum transport must have 
occurred. This is seen in merger simulations, but
is contrary to alternative models of episodic formation histories 
without mergers.

\vspace{-6pt}
\section{Conclusions}

\vskip -2pt
	The merger model of Ashman \& Zepf (1992) made four predictions
for properties of globular cluster systems that had not yet been tested 
by observation, and which were generally contrary to the standard
picture at that time. Three of
these four have now been confirmed by observation, while the fourth
has yet to be tested. Although limitations of the original model
have also been revealed by other observations, the striking agreement
with many of the predictions suggests that the model is mostly correct.
For it not to be correct, the young clusters observed in mergers
would have to be irrelevant for globular cluster populations as 
a whole, even though their properties are exactly those expected
for young globular clusters.
Moreover, the bimodal metallicity distributions observed in elliptical 
galaxy GCSs would have to have formed in an episodic process other than
mergers, that is also efficient at transporting angular momentum.
Given these constraints, any model that successfully 
accounts  for the observations is likely to be similar physically 
to the merger model. Globular clusters have greatly improved
our understanding of the formation history of galaxies,
and with further photometric observations and new 
spectroscopic data, they promise to continue to do so.

\centerline{\bf Acknowledgements}

	We thank our many collaborators on the various projects described
above. Some of the research described above was supported by grants
GO-06092.01-94A, AR-06405.01-95A, and an award to SEZ from the Dudley 
Foundation. SEZ also acknowledges the support of a Hubble Fellowship during
much of this work.


\end{document}